\documentclass[aps,pra,reprint,nofootinbib,twocolumn,superscriptaddress,showpacs,showkeys,longbibliography]{revtex4-1}
\usepackage{eurosym}
\usepackage{amsmath,amssymb,amstext}
\usepackage[usenames,dvipsnames]{color}
\usepackage{graphicx}
\usepackage{braket}
\usepackage{natbib}
\usepackage{comment}
\usepackage{dcolumn}
\usepackage[english]{babel}
\usepackage{wasysym}
\usepackage[colorlinks,bookmarks=false,citecolor=blue,linkcolor=red,urlcolor=blue]{hyperref}

\begin{document}
\title{Dark-bright solitons in spinor polariton condensates under nonresonant pumping}
\author{Xingran Xu}
\affiliation{Shenyang National Laboratory for Materials Science, Institute of Metal Research, Chinese Academy of Sciences, Shenyang, China}
\affiliation{ School of Materials Science and Engineering, University of Science and Technology of China, Hefei, China}
\affiliation{Department of Physics, Zhejiang Normal University, Jinhua, 321004, China}
\author{Lei Chen}
\affiliation{School of Physics and Electronic Science, Zunyi Normal University, Zunyi 563006, China}
\author{Zhidong Zhang}
\affiliation{Shenyang National Laboratory for Materials Science, Institute of Metal Research, Chinese Academy of Sciences, Shenyang, China}
\affiliation{ School of Materials Science and Engineering, University of Science and Technology of China, Hefei, China}
\author{Zhaoxin Liang}
\email{The corresponding author: zhxliang@gmail.com}
\affiliation{Department of Physics, Zhejiang Normal University, Jinhua, 321004, China}
\date{\today}
\begin{abstract}

Adopting a mean-field Gross-Pitaevskii description for a spinor polariton Bose-Einstein condensates under non-resonant pumping, we investigate the static and dynamical properties of dark-bright solitons. We derive analytically the equation of motion for the center of mass of the dark-bright soliton center, using the Hamiltonian approach. The resulting equation captures how the combination of the open-dissipative character and the spin degrees of freedom of a polariton Bose-Einstein condensate affects the properties of a dark-bright soliton, i.e. the dark-bright soliton relaxes by blending with the background at a finite time. In this case, we also determine the life time of the DB soliton. Further numerical solutions of the modified dissipative two-component Gross-Pitaevskii equations are in excellent agreement with the analytical results. In presence of the Langevin noise, we demonstrate that the DB solitons can still propagate for a long time, which is sufficient for their experimental observations within current facilities.

 \end{abstract}
\maketitle

\section{Introduction}\label{section:one}

Soliton~\cite{Kevrekidis2015} is a solitary wave packet which can maintain its shape due to a self-stabilization against dispersion through nonlinear interaction~\cite{SolitonReview0,SolitonReview1,SolitonReview2,SolitonReview3,SolitonReview4}. In one-component nonlinear Schr\"odinger equation (NLSE), 
the dark~\cite{DarkReviewKivshar1988,Dark1} (bright~\cite{SolitonReview4,Liang2005}) soliton can exist provided the interaction is repulsive (attractive). By contrast, in two-component (spinor) NLSE with repulsive interaction, i.e., the vector variant of NLSE, a dark-bright (DB) soliton~\cite{DBsoliton1} can appear even though a bright soliton alone is forbidden: A density dip in the form of a dark soliton in one-component plays the role of an effective potential well for a bright soliton created in the second component~\cite{DBsoliton1}. This remarkable phenomenon highlights the important role of the spin degree of freedom (two-component) of system, which, when interplaying with the dispersion and nonlinearity, can give rise to novel solitons with no analogue in the one-component counterpart~\cite{DBTheo1,DBTheo2,DBTheo3,DBTheo4,DBTheo5}. DB solitons have been extensively studied in a wide variety of physical systems. In particular, they have been experimentally observed in the context of nonlinear optics~\cite{DBexpOpt1,DBexpOpt2} and recently in atomic condensates~\cite{DBexpBEC1,DBexpBEC2,DBexpBEC21,DBexpBEC3}.
 
The recent realization of Bose-Einstein condensates (BECs) of polaritons~\cite{Rev1,Rev2,Rev3} in quantum-well semiconductor microcavities has opened intriguing possibilities to explore the DB solitons~\textit{beyond thermal equilibrium}. As polaritons undergo rapid radiative decay, their population in the condensate is maintained by persistent optical pumping. Hence a polariton condensate is inherently non-equilibrium with open-dissipative character~\cite{Rev2}. Its mean-field physics can be well captured by a Gross-Pitaevskii equation (GPE) with gain and loss ~\cite{Wouters2007,Xu2017}, where the nonlinearity results from the strongly and repulsively interacting excitons~\cite{Rev1}. Further, polaritons naturally possess peculiar spin properties ~\cite{Rev0}: Due to the strong couplings between excitons and photons, the spin-up and spin-down projections of the total angular momentum of excitons along the growth axis of the structure directly come from the right- and left-circularly polarized photons absorbed or emitted by the cavity, respectively. Combinations of these exceptional properties promise polariton BECs as a novel platform for realization and investigation of
various nonlinear phenomena, such as solitons.

In the one-component polariton condensates formed under resonant pumping laser, a series of experiments have demonstrated the existence of the oblique dark solitons and vortices~\cite{Amo2011,Grosso2011,Grosso2012}, or the bright spatial and temporal solitons~\cite{Sich2011,Ostrovskaya2013}. In condensates created spontaneously under incoherently pumping, the formation and behavior of dark solitons have been investigated in Refs.~\cite{Xue2014,Smirnov2014,Xue2018} theoretically and observed~\cite{DarkSolitonNExp} experimentally. In the non-resonantly driven spinor polariton BEC at one dimension, existence of stable dark soliton trains has been reported~\cite{Pinsker2014}. Yet, the existence of DB solitons in the non-equilibrium polariton BEC remains unexplored. It is the aim of the present work to investigate the interplay of the nonlinearity, dispersion and the spin degree of freedom of system on the DB solitons in the presence of incoherent driving and dissipation.

In this work, we study the physics of DB solitons in a spinor polariton BEC formed under non-resonant pumping, by solving the two-component dissipative GPEs with a combination of analytical and numerical approaches. Our goal is to explore the combined effects of the open-dissipative and spinor nature on the dynamics of DB solitons. To this end, we first use the Hamiltonian approach and analytically derive the evolution equations for the soliton parameters, i.e. the inverse width of the soliton. We compare this analytical result with the numerically solutions for the trajectory of DB solitons directly obtained from the GPEs, and find a remarkable agreement between the two. Further, we have demonstrated, while noise will eventually destroy the dissipative DB solitons, their life time remains sufficient for a feasible experimental observation. Our results open a route to observe stable DB solitons in the non-equilibrium spinor polariotn BEC within the current experimental apabilities.

The paper is organized as follows. In Sec.~\ref{section:two}, we introduce the two-component dissipative GPEs coupled to the rate equation of a spin-unpolarized reservoir, which can well describe the static and dynamical properties of the polariton BECs under nonresonant pumping. In Sec.~\ref{Stability}, we present a
brief review on the collective excitations in the uniform stationary state using the Bogoliubov's theory, and provide analytical
results for the parameter regimes where the system is modulationally stable. In Sec. In Sec.~\ref{Dynamics}, we use the Hamiltonian approach to derive the equation of motion for the center of mass of the DB solitons. We then compare the analytical results with the exact numerical ones. We moreover demonstrate how these dissipative solitons can be affected by the Langevin noise by numerically solving the stochastic GPEs. In Sec.~\ref{Conclusion}, we conclude with a summary of our main results and final remarks.

\section{Model System}\label{section:two}

We consider a spinor exciton-polariton BEC created under
nonresonant pumping~\cite{Rev0,Xu2017}. At the mean-field level, the condensate can be well described
by a two-component time-dependent wave function $[\psi_1,\psi_2]$. For the excitonic reservoir, we assume that the spin relaxation of the reservoir is sufficiently
fast so that the reservoir on the relevant time scales can be modeled by a scalar density denoted by $n_R(t)$~\cite{Rev0,Rev1}. 

The dynamics of the spinor polariton BEC can be described by the following driven-dissipative two-component GPEs~\cite{Wouters2007,Xu2017,Borgh2010,Liew2015,Li2015,Askitopoulos2016,Rev1}
\begin{eqnarray}
i\hbar\frac{\partial \psi_1}{\partial t}&=&\left[-\frac{\hbar^2\nabla^2}{2m}+g\left|\psi_1\right|^2+g_{12}\left|\psi_2\right|^2+g_Rn_R\right]\psi_1\nonumber\\
&+&\frac{i\hbar}{2}\left(R n_R-\gamma_C \right)\psi_1+i\hbar\frac{d\psi^{st}_1}{dt},\label{SpinorP1}\\
i\hbar\frac{\partial \psi_2}{\partial t}&=&\left[-\frac{\hbar^2\nabla^2}{2m}+g\left|\psi_2\right|^2+g_{12}\left|\psi_1\right|^2+g_Rn_R\right]\psi_2\nonumber\\
&+&\frac{i\hbar}{2}\left(R n_R-\gamma_C\right)\psi_2+i\hbar\frac{d\psi^{st}_2}{dt}.\label{SpinorP2}
\end{eqnarray}
Here,  $m$ is the mass of polariton, $g$ ($g_{12}$) is the interaction between polaritons with same (opposite) spins, and $g_R$ is the interaction strength between reservoir and polariton. Further,
the terms $d\psi_i^{st}=2D_id W_i$ ($i=1,2$) account for the fluctuations induced by a white noise~\cite{Noise1,Noise2,Noise3}. Here $d W_i$ is a Gaussian random variable characterized
by the correlation functions~\cite{Noise1,Noise2,Noise3} 
\begin{eqnarray}
\langle d W^*_id W_j\rangle&=&\delta_{i,j}dt, \nonumber \\
\langle d W_id W_j\rangle&=&0,\label{Noise}
\end{eqnarray}
 where $i$, $j$ are indices for the component of spinor polariton BEC. Below we will ignore the noise term in Eq.~(\ref{Noise}) when calculating the stationary states, analyzing the modulation stability parameter regimes and deriving the equation of motion for  DB soliton's width. We will add the noise terms afterwards which serves to test the stability of DB solitons against fluctuations. Furthermore, in Eqs.~(\ref{SpinorP1}) and (\ref{SpinorP2}) we have ignored the effects of the transverse-electric and transverse-magnetic splitting~\cite{Shelykh2004,Shelykh2006}. Note that going beyond the Gross-Pitaevskii equations (\ref{SpinorP1}) and (\ref{SpinorP2}) to fully include the quantum and thermal fluctuations of the quantum field (e.g., Keldysh path-integral method \cite{ByPI1,ByPI2,ByPI0,ByPI01}) is beyond the scope of this work. 

We consider Eqs.~(\ref{SpinorP1}) and (\ref{SpinorP2}) are coupled to a (scalar) incoherent reservoir as mentioned earlier, which is described by a rate equation, i.e.,
\begin{eqnarray}
\frac{\partial n_R}{\partial t}=P-\gamma_R n_R-R\left(\left|\psi_1\right|^2+\left|\psi_2\right|^2\right)n_R.\label{Rate}
\end{eqnarray}
Here, $P$ is the rate of an off-resonant continuous-wave (cw)
pumping, $\gamma_R^{-1}$ describes the lifetime of reservoir polaritons,
and $R$ is the stimulated scattering rate of reservoir polaritons
into the spinor condensate.

In our subsequent analytical treatment, we will consider the situation when $g>0$ ~\cite{Rev0, Ciuti98} and $g_{12}>0$ with $g=g_{12}$. We note that while in present experiments one typically has $g_{12}<0$ with $|g_{12}|\ll g$, recent experimental progress in realizing tunable cross-spin interaction properties has opened prospect into realizing polariton BECs in regimes $g_{12}>0$ and $g_{12}/g\ge 1$~\cite{NP2014,Feshbach2017}. Under these conditions, in the steady state, the spinor polariton BEC has a total condensate density $n_0=(P-P_{\text {th}})/\gamma_C$, which is equally distributed in each component, while the reservoir density is given by 
$n_R^0=\gamma_C/R$.

For convenience, we will recast equations (\ref{SpinorP1}-\ref{Rate}) into a dimensionless form. After rescaling $\psi\rightarrow \psi/\sqrt{n_0}$ and introducing $m_R=n_R-n_R^{0}$, we obtain
\begin{eqnarray}
i\frac{\partial}{\partial t}\psi_{1}&=&-\frac{1}{2}\nabla^{2}\psi_{1}+\left( \left|\psi_{1} \right|^{2}+\left|\psi_{2} \right|^{2}-1 \right)\psi_{1}\nonumber\\ &+&\bar{{g}}_{R}m_{R}\psi_{1}+\frac{i}{2}{\bar{R}}m_{R}\psi_{1}+i\frac{d\psi^{st}_1}{dt},\label{s1}\\
i\frac{\partial}{\partial t}\psi_{2}&=&-\frac{1}{2}\nabla^{2}\psi_{2}+\left( \left|\psi_{1} \right|^{2}+\left|\psi_{2} \right|^{2}-\bar{\mu} \right)\psi_{2}\nonumber\\ &+&\bar{{g}}_{R}m_{R}\psi_{2}+\frac{i}{2}{\bar{R}}m_{R}\psi_{2}+i\frac{d\psi^{st}_2}{dt},\label{s2}\\
\frac{\partial}{\partial t}m_{R}&=&\bar{\gamma}_{C}\left(1-\left|\psi_{2}\right |^{2}-\left|\psi_{1}\right|^{2}\right)-\bar{\gamma}_{R}m_{R}\nonumber\\
&-&\bar{R}\left(\left|\psi_{1}\right|^{2}+\left|\psi_{2}\right|^{2}\right)m_{R},\label{RD}
\end{eqnarray}
Here we have introduced the notations
$\bar{P}=\hbar P/\left(gn_{0}^{2}\right)$, $\bar{\mu}\rightarrow\mu_{2}/\mu_{1}$, $\bar{R}=R/gn_0$, and $\bar{\gamma}_{C}=\gamma_{C}/gn_{0}$. Moreover, the time $t$ and space coordinate $x$ are measured in the units of $\hbar/gn_0$ and $\xi=\sqrt{\hbar^2/mgn_0}$, respectively.

Our goal next is to develop a theory for the propagation and stability of DB solitons of a spinor polariton BEC under incoherent uniform pumping, based on solutions of Eqs.~(\ref{s1})-(\ref{RD}). Before proceeding, we comment that a dark soliton represents a nonlinear collective excitation with finite amplitude, which is generated out of a homogeneous condensate. Therefore, as a first step, it is important to check whether the homogenous background itself is stable against weak perturbations, which is the aim of the next section.

\section{Stability of a homogeneous condensate}\label{Stability}

Here we briefly revisit the known results on the linear collective excitations of a spinor polariton BEC under an incoherent uniform pumping. We will determine analytically the regime of modulational instability (MI) \cite{ModulationInstability} of a homogeneous background.

Following standard procedures of stability analysis \cite{Xu2017}, we linearize Eqs.~(\ref{s1}) - (\ref{RD}) around the steady-state solution $(1/\sqrt{2},1/\sqrt{2},0)^T$ and obtain the eigenvalue problem for the dispersion relation, i.e., 
\begin{eqnarray}
\left(\omega^{2}-k^4/4\right)\times\Big[\omega^{3}&+&i\left(\bar{\gamma}_{R}+\bar{R}\right)\omega^{2}
-\left(\omega_{B}^{2}
+\bar{R}\bar{\gamma}_{C}\right)\omega\nonumber\\
&-&i\omega_{B}^{2}\left(\bar{\gamma}_{R}+\bar{R}\right)+i\bar{\gamma}_{C}\bar{g}_{R}k^{2}\Big]=0, \label{Phase1General}
\end{eqnarray}
with $\omega^2_B=k^2+k^4/4$.

We now determine the MI regime from Eq.~({\ref{Phase1General}). Specifically, there exist five complex dispersion branches $\omega_j(k)=\Re(\omega_j(k))+i\Im(\omega_j(k)) $, ($j=1,2,3,4,5$). The modulation instability occurs when $\Im(\omega_j)>0$ at certain momenta. In our case,  MI is found to occur under the condition of $\bar{g}_R\bar{\gamma}_C>\bar{\gamma}_R+\bar{R}$, i. e.  in a regime $k\in [k_1,k_2]$ with $k_1=0$ and 
\begin{equation}
k_2=2\sqrt{\frac{\bar{g}_R\bar{\gamma}_C}{\bar{\gamma}_R+\bar{R}}-1}.
\end{equation}
In the remaining part of our work, we will restrict our consideration in the fast reservoir limit $\gamma_R\gg \gamma_C$, where the modulation stability condition is naturally satisfied. 

\section{Dynamics of dark-bright soliton}\label{Dynamics}

In the limit of fast reservoir, we derive in this section the equation of motion for the center of mass of the DB soliton supported by Eqs.~(\ref{s1})-(\ref{RD}) in the absence of noise. We find it is more transparent to rewrite Eqs.~(\ref{s1})-(\ref{s2}) into the following form
\begin{eqnarray}
i\frac{\partial}{\partial t}\psi_{1}+\frac{1}{2}\nabla^{2}\psi_{1}-\left( \left|\psi_{1} \right|^{2}+\left|\psi_{2} \right|^{2}-1 \right)\psi_{1}&=&R_1,\label{sss1}\\
i\frac{\partial}{\partial t}\psi_{2}+\frac{1}{2}\nabla^{2}\psi_{2}-\left( \left|\psi_{1} \right|^{2}+\left|\psi_{2} \right|^{2}-\bar{\mu} \right)\psi_{2}&=&R_2,\label{sss2}
\end{eqnarray}
where we have introduced the notation
\begin{equation}
R_{1,2}=\left(\bar{g}_{R}m_{R}+\frac{i}{2}\bar{R}m_{R}\right)\psi_{1,2}.\label{Pert}
\end{equation}
The dynamics of the perturbation $m_R$ in Eq.~(\ref{Pert}) is governed by the rate equation
\begin{eqnarray}
\frac{\partial}{\partial t}m_{R}&=&\bar{\gamma}_{C}\left(1-\left|\psi_{2}\right |^{2}-\left|\psi_{1}\right|^{2}\right)-\bar{\gamma}_{R}m_{R}\nonumber\\
&-&\bar{R}\left(\left|\psi_{1}\right|^{2}+\left|\psi_{2}\right|^{2}\right)m_{R}.\label{PertRD}
\end{eqnarray}
Equations (\ref{sss1}) and (\ref{sss2}) can be viewed as two coupled NLSEs subjected to time-dependent perturbations provided by $R_1$ and $R_2$. (We note that the two equations obviously reduce to that in Ref.~\cite{Smirnov2014} if one of the component vanishes.) 

As a first step, we consider $R_{1,2}=0$, i.e., in the absence of the open-dissipative effects. Under the boundary conditions $\psi_1\rightarrow 1$ and $\psi_2\rightarrow 0$ as $|x|\rightarrow \infty$, there exists an exact one-DB-soliton solution, which takes the form
\begin{eqnarray}
\psi_{D}&=&\cos\phi \tanh\left[D(x-x_{0}(t))\right]+i \sin\phi, \label{Darksoliton}\\
\psi_{B}&=&\eta \text{sech}\left[D(x-x_0(t))\right]e^{ik[x-x_0(t)]+i\theta\left(t\right)}. \label{Brightsoliton}
\end{eqnarray}
Here $\phi$ is the dark soliton's phase angle, $\cos\phi$ and $\eta$ are amplitudes  of the dark and bright solitons, $D$ and $x_0(t)$ are associated with the inverse width and the center position of the dark and bright solitons. Further, $k=D\tan\phi$ is referred to as the wavenumber of the bright soliton. The amplitude of bright soliton $\eta$ is connected to the number $n_2$ of atoms in the bright soliton through the following relation~\cite{Smirnov2014}
\begin{equation}
\int_{-\infty}^{\infty}\left|\psi_{2}\right|^{2}dx=\frac{2\eta^{2}}{D}=n_2\label{eta}.
\end{equation}
Notice that the above parameters of the DB solitons are not independent, related to each other by $D^2=\cos^2\phi-\eta^2$, $\dot{x}_0=D\tan\phi$, and $\theta\left(t\right)=\left(D^{2}-k^{2}\right)t/2+(\bar{\mu}-1)t$.

Next, we account for the open-dissipative effects as captured by $R_{1,2}\neq 0$. We will rely on the Hamiltonian approach~\cite{Perturbation1,Smirnov2014} of the perturbation theory which allows for analytical treatment of the effect of $R_{1,2}$ [see (\ref{Pert})] on the DB solitons. We note that, different from Ref.~\cite{Smirnov2014}, the system considered here has spin degrees of freedom and is described by two coupled GP equations. At the heart of the Hamiltonian approach of quantum dynamics for DB solitons is the assumption that, in presence of perturbation, the parameters of the solitons become slow functions of time, but the functional
form of the soliton remains unchanged, i.e., $\phi\rightarrow \phi(t)$ and $D\rightarrow D(t)$, while keeping $D^2(t)=\cos^2\phi(t)-\eta^2$, $\dot{x}_0(t)=D(t)\tan\phi(t)$.
As such, the time evolution of the parameters $\phi(t)$ and $D(t)$ can be obtained from the time evolution of the DB soliton energy \cite{Smirnov2014}, i.e.,
\begin{equation}
\frac{dE}{dt}=-2\Re\int \left[R_{1}\frac{\partial}{\partial t}\psi_{1}^{*}+R_{2}\frac{\partial}{\partial t}\psi_{2}^{*}\right]dx,\label{DBevolution}
\end{equation}
where $E$ is the energy of the DB soliton given by 
\begin{eqnarray}
E=\frac{1}{2}\int dx\big\{\left|\nabla\psi_{1}\right|^{2}+\left |\nabla\psi_{2}\right|^{2}+(\left|\psi_{1}\right|^{2}+\left|\psi_{2}\right|^{2}-1)^{2}\big\}.\label{energy}
\end{eqnarray}
We remark that Eqs.~(\ref{DBevolution}) and (\ref{energy}) can be regarded as
the so-called adiabatic invariant approximation of the perturbation theory of solitons in Refs.~\cite{AdiabaticMethod1,AdiabaticMethod2}. We also stress that
our theory so far assumes a perturbative regime of reservoir excitations $m_R(\vec{r},t)\ll 1$.

Substituting Eqs.~(\ref{Darksoliton}) and (\ref{Brightsoliton}) into Eq.~(\ref{energy}), we obtain the time-dependent energy of the DB soliton as follows
\begin{eqnarray}
E=\frac{4}{3}D^{3}(t)+\frac{n_{2}}{2}D^{2}(t)\sec^{2}\phi(t).\label{DBenergy}
\end{eqnarray}
The first (second) term in Eq.~(\ref{DBenergy}) corresponds to the dark (bright) soliton's energy, respectively. For $n_2=0$, i.e., when bright soliton vanishes, Eq.~(\ref{DBenergy})
reduces to the energy of a single-dark soliton [see Eq.~(30) in Ref.~\cite{Smirnov2014}]. However, when the dark soliton disappears, the bright soliton will vanish as well. To see this, note that the darkness of the dark soliton in the DB solitons can be approximately measured by the velocity $\sin \phi$ via the relation $n^{\text{min}}_D(t)=\sin^2\phi$. Hence, when the dark soliton vanishes for $n^{\text{min}}_D\rightarrow 1$ [i.e. when $\phi=\pi/2$], the second term in Eq.~(\ref{DBenergy}) becomes divergent, meaning the bright soliton will be destroyed. This provides an intuitive understanding that a DB soliton can exist in the sense that a density dip in the form of a dark soliton in one-component creates a potential well for the second component which traps a bright soliton therein. It then follows from Eq.~(\ref{DBenergy}) that the time variation of DB soliton's energy is described by
\begin{equation}
\frac{dE}{dt}=4D^{2}\dot{D}+n_{2}D\sec^{2}\phi\left(\dot{D}+D \tan\phi\dot{\phi}\right).\label{PertLeft}
\end{equation}

Next, calculations of the right side of Eq.~(\ref{DBevolution}) require the knowledge of the reservoir density $m_R$. In the limit of fast reservoir, we have
\begin{eqnarray}
\bar{m}_R=\frac{\bar{\gamma}_{C}}{\bar{\gamma}_{R}}\left(1-\left|\psi_{1}\right|^{2}-\left|\psi_{2}\right|^{2}\right).\label{MR0}
\end{eqnarray}
Using Eqs.~(\ref{Pert}) and (\ref{MR0}), we write the right side of Eq.~(\ref{DBevolution}) as
\begin{eqnarray}
&&\Re\int \left(R_{1}\frac{\partial}{\partial t}\psi_{1}^{*}+R_{2}\frac{\partial}{\partial t}\psi_{2}^{*}\right)dx
\nonumber\\
&=&\frac{2\bar{R}}{3}\frac{\bar{\gamma}_{C}}{\bar{\gamma}_{R}}D^{3}\sin^{2}\phi+\frac{\bar{g}_{R}}{3}\frac{\bar{\gamma}_{C}}{\bar{\gamma}_{R}}\cos^{2}\phi\left(\dot{D}+4D\tan\phi\dot{\phi}\right)\nonumber\\
&-&n_{2}\frac{\bar{R}}{3}\frac{\bar{\gamma}_{C}}{\bar{\gamma}_{R}}D^{4}\tan^{2}\phi+n_{2}\frac{\bar{g}_{R}}{6}\frac{\bar{\gamma}_{C}}{\bar{\gamma}_{R}}D\dot{D}.\label{PertRight}
\end{eqnarray}
In Eq.~(\ref{PertRight}), the terms in the second (third) lines result from the coupling of dark (bright) solitons with the reservoir. When the bright soliton vanishes ($n_2=0$), the first term in the second line exactly recovers the corresponding result in Ref.~\cite{Smirnov2014}. Whereas, the second term in the second line comes from the reservoir-induced modification of the interaction of polaritons, which effect has been ignored in Ref.~\cite{Smirnov2014}.

\begin{figure*}
  \includegraphics[width=0.8\textwidth]{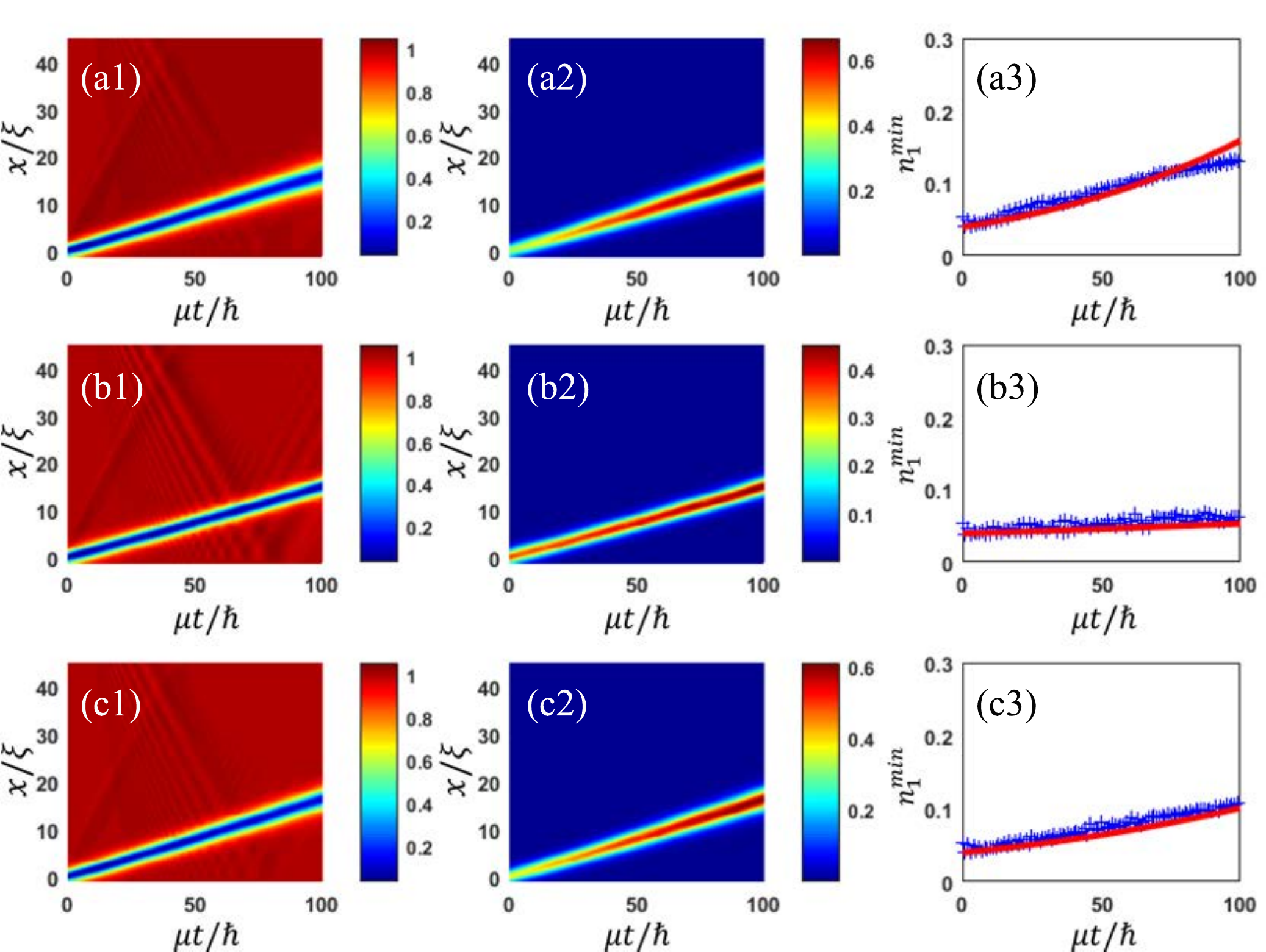}\\
\caption{Dynamics of the one-dimensional DB soliton of a spinor polariton BEC created under nonresonant pumping. Shown are the contour plots of  the dark soliton of $\left|\psi_D\right|^2$ (left column),  the bright soliton of $\left|\psi_B\right|^2$ (middle column), and  the dependence of the darkness $n_1^{\text {min}}$ by numerically solving Eqs.~(\ref{sss1})-(\ref{sss2}) (right column, blue plus signs). The solid red lines in (c1)-(c3) are calculated using the analytical results in Eq.~(\ref{Darkness}) in the maintext. In all plots, we have chosen $\bar{g}_R$=2, $\bar{\gamma}_C$=3, $\phi_0$=0.2. For other parameters, (a1)-(a3): $\bar{R}=0.5$, $\bar{\gamma}_R$=30;  (b1)-(b3):  $\bar{R}$=0.1; $\bar{\gamma}_R$=30; and (c1)-(c3): $\bar{R}$=0.5, $\bar{\gamma}_R=45$.}\label{Fig1}
\end{figure*}

Finally, substitutions of Eqs.~(\ref{PertLeft}) and (\ref{PertRight}) into Eq.~(\ref{DBevolution}}) readily yields the equation of motion for the inverse width of soliton $D$, i.e., 
\begin{widetext}
\begin{eqnarray}
\left[1-\frac{\bar{g}_{R}}{2}\frac{\bar{\gamma}_{C}}{\bar{\gamma}_{R}}+\frac{n_{2}^{2}}{4D^2\left(2D+n_{2}\right)^{2}}\right]\dot{D}&=&-\frac{1}{3}\bar{R}\frac{\bar{\gamma}_{C}}{\bar{\gamma}_{R}}D\left[\left(1-D^{2}\right)-\frac{n_2}{2D+n_{2}}\right].\label{EOM}
\end{eqnarray}
\end{widetext}
Equation (\ref{EOM}) for $n_2=0$ reduces to the well-known results in Ref.~\cite{Smirnov2014}. In this case, Eq.~(\ref{EOM}) can be simplified into $\left(1-\frac{\bar{g}_{R}}{2}\frac{\bar{\gamma}_{C}}{\bar{\gamma}_{R}}\right)\dot{D}=-\frac{1}{3}\bar{R}\frac{\bar{\gamma}_{C}}{\bar{\gamma}_{R}}D\left(1-D^{2}\right)$. In terms of $D=\sqrt{1-v_{\text{s}}^2}$ as in Ref.~\cite{Smirnov2014},  we have $\frac{dv_{s}}{dt}=\frac{1}{2\tau}\left(1-v_{s}^{2}\right)v_{s}$ with $\tau=\frac{3}{2}\left(1-\frac{\bar{g}_{R}}{2}\frac{\bar{\gamma}_{C}}{\bar{\gamma}_{R}}\right)\frac{\bar{\gamma}_{R}}{\bar{R}\bar{\gamma}_{C}}$. Here, for vanishing $g_R$,  our result can exactly recover the previous result, i.e., Eq.~(33) in Ref.~\cite{Smirnov2014}.  However, different from Ref.~\cite{Smirnov2014}, our calculations have considered the interactions between the condensate and the reservoir characterized by $g_R$. The presence of such coupling will modify the effective local self-induced potential exhibited by the condensate, and therefore, changes the width and darkness of the solitonic state.

Equation (\ref{EOM}) with $n_2\neq 0$ allows us to analyze the combined effects of the spinor and open-dissipative nature of polariton BEC on the dynamics of the DB solitons. Notice that the velocity of the dark soliton is defined as $v_{\text{s}}=D\tan\phi$. By virtue of the relation $v_{s}=\sqrt{(2D-2D^{3}-n_{2}D^{2})/(2D-n_{2})}$, Equation (\ref{EOM}) can be written as
\begin{equation}
m_{\text{eff}}\frac{dv_{s}}{dt}=F_{\text{eff}}\left(v_{s}\right).\label{EOMV}
\end{equation}
Here $m_{\text{eff}}$ plays the role of effective mass of the DB solitons 
\begin{equation}
m_{\text{eff}}=1-\frac{\bar{g}_{R}}{2}\frac{\bar{\gamma}_{C}}{\bar{\gamma}_{R}}+\frac{n_{2}^{2}}{4D^{2}\left(2D+n_{2}\right)^{2}},\label{EM}
\end{equation}
and $F_{\text{eff}}$ represents an effective force given by 
\begin{eqnarray}
F_{\text{eff}}&=&\frac{1}{3}\bar{R}\frac{\bar{\gamma}_{C}}{\bar{\gamma}_{R}}D\left[\left(1-D^{2}\right)-\frac{n_{2}}{2D+n_{2}}\right]\nonumber\\
&\times&\frac{D\left[2D\left(2D-n_{2}\right)-n_{2}^{2}\right]+n_{2}}{\left(2D-n_{2}\right)\sqrt{D\left[2-D\left(2D+n_{2}\right)\right]\left(2D-n_{2}\right)}}.\label{Veff}
\end{eqnarray}

Equation (\ref{EOMV}) allows us to interpret the dynamics of the DB solitons in terms of the motion of a classical particle of mass $m_{\text{eff}}$ subjected to an external force $F_{\text{eff}}$. Equation (\ref{EM}) trivially recovers the corresponding result in Ref.~\cite{Smirnov2014} when $n_2=0$. Since the dark and bright solitons in a DB soliton share the same velocity, we will below focus on analyzing the behavior of the dark one. An interpretation of Eq.~(\ref{EOMV}) is straightforward following from the Eq.~(\ref{sss1}) for the dark soliton wavefunction $\psi_1$, which can be rewritten as 
\begin{eqnarray}
i\frac{\partial \psi_{1}}{\partial t}+\frac{\Delta}{2}\psi_{1}+g_{e}\left(1-\left|\psi_{1}\right|^{2}\right)\psi_{1}-V\psi_{1}&=iP.\label{EOMeffective}
\end{eqnarray}
Here $g_{\text{e}}=1-\bar{g}_{R}\bar{\gamma}_{C}/\bar{\gamma}_{R}$ is the effective interaction constant, $V=\left(1-\bar{g}_{R}\bar{\gamma}_{C}/\bar{\gamma}_{R}\right)\left|\psi_{2}\right|^{2}$ is the effective external potential induced by the bright soliton and $P=\frac{1}{2}\bar{R}\bar{\gamma}_{C}/\bar{\gamma}_{R}\left(1-\left|\psi_{1}\right|^{2}-\left|\psi_{2}\right|^{2}\right)\psi_{1}$. Based on Eq.~(\ref{EOMeffective}), the equation of motion for the center of mass of the dark soliton's in Eq.~(\ref{EOMV}) can be explained as follows:

(i) The effective mass in Eq.~(\ref{EM}) involves two corrections due to the coupling of polariton BEC to reservoir (see second and third terms). To understand the first modification, we note that according to Eq.~(\ref{EOMeffective}), the interaction between the dark soliton and the reservoir, characterized by $\bar{g}_R$, modifies the effective local self-induced potential exhibited by the condensate [see the term containing $g_e$ in Eq.~(\ref{EOMeffective})]. This effect changes both the width and darkness of the solitonic state, and therefore, the first correction. On the other hand, the coupling of the bright soliton and reservoir results in an effective external potential [see the term of $V$ in Eq.~(\ref{EOMeffective})], hence explaining the second correction to the effective mass in Eq.~(\ref{EM}).

(ii) To understand the effective force in Eq.~(\ref{Veff}), we note that the stimulated scattering term proportional to $\bar{R}$ on the right side of Eq.~(\ref{EOMeffective}) is responsible for the variation of the dark soliton's velocity. The effective force in Eq.~(\ref{Veff}) contains two competitive parts: While the action of the first part leads to acceleration of the dark soliton, the second part - which is induced by the bight soliton - will slow down the motion of the dark soliton.

\begin{figure*}
  \includegraphics[width=0.8\textwidth]{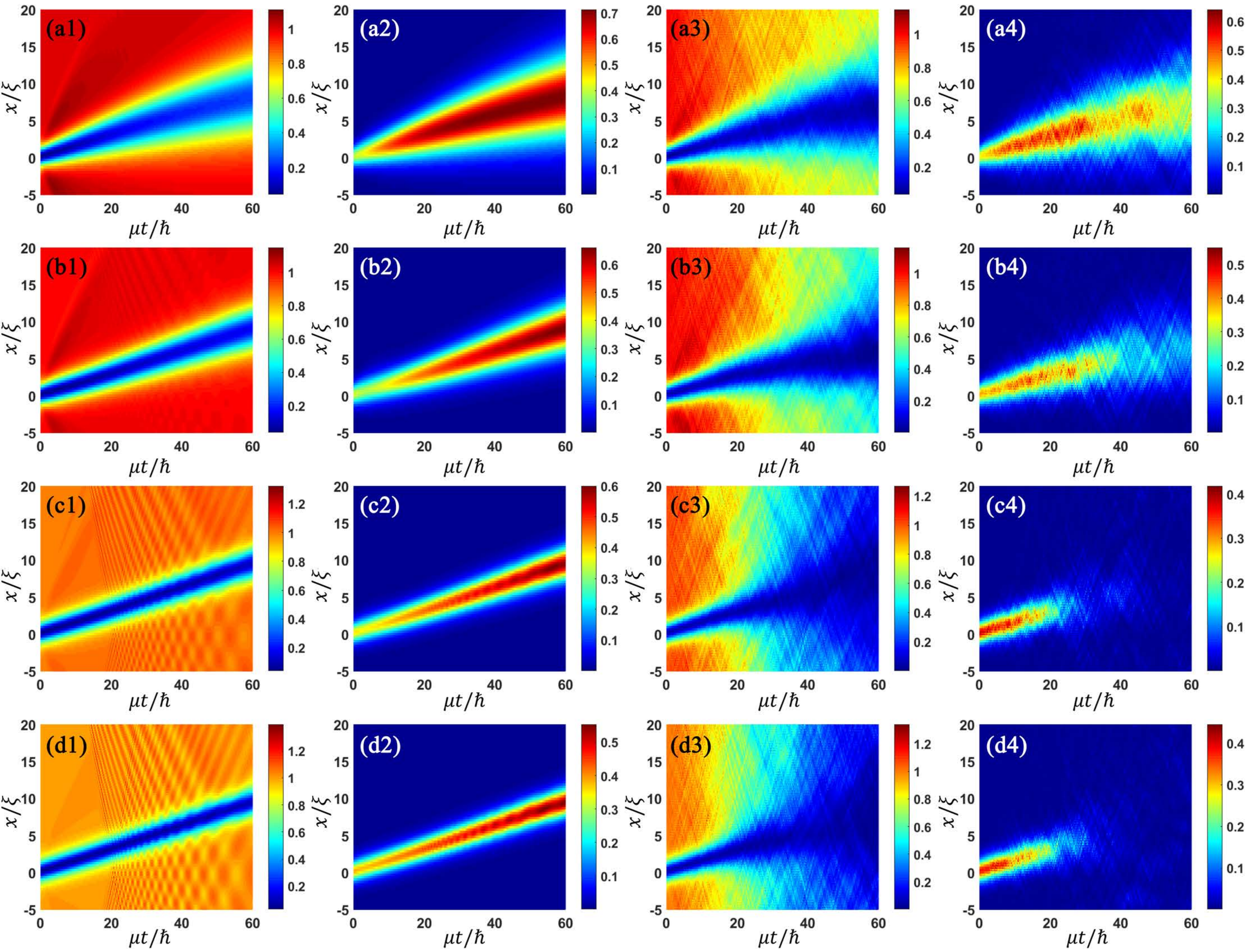}\\
\caption{Effects of Langevin noise on the dynamics of DB solitons in a spinor polariton BEC under nonresonant pumping by numerically solving Eqs.~(\ref{s1})-(\ref{RD}).  As a comparison, the first and second columns depict the time evolution of DB solitons without Langevin noise, while the third and forth columns correspond to the presence of Langevin noise [see Eq.~(\ref{Noise})] with $d\psi_i^{st}$=0.01$dW_i$. In all plots, we take $\bar{g}_R$=2, $\bar{\gamma}_C$=3, $\phi$=0.2. For other parameters, (a1)-(a4): $\bar{R}$=1.5, $\bar{\gamma}_R$=15; (b1)-b4: $\bar{R}=0.5$, $\bar{\gamma}_R$=15;  (c1)-(c4): $\bar{\gamma}_C$=0.5, $\bar{\gamma}_R$=30; (d1)-(d4): $\bar{R}=0.5$, $\bar{\gamma}_R$=45.  }\label{Fig2}
\end{figure*}

We now proceed to solve the dynamics of DB solitons governed by Eq.~(\ref{EOM}). Searching for the equilibrium solution, we set $\dot{D}=0$ in Eq.~(\ref{EOM}), i.e. $\left(1-D^{2}\right)-\frac{n_{2}}{2D+n_{2}}=0$, which yields $\phi_{eq}=0$, $x_{0eq}=0$, and $D_{eq}=-n_{2}/4+\sqrt{n_{2}^2/16+1}$. Next, we expand the solutions around the equilibrium values, i.e., $x_0(t)\rightarrow 0+x_0(t)$, $\phi(t)\rightarrow 0+\phi(t)$, $D(t)\rightarrow D_{eq}+D_1(t)$. We recall that the three parameters are not independent and are related to each other as earlier described. Therefore, as $\phi(t)\approx0$ near the equilibrium value, we have $D_1(t)=-\tilde{D}\phi^2(t)$ with $\tilde{D}=(2D_{eq}+n_{2}/2)^{-1}$. With these approximations, Equation (\ref{EOM}) can be readily calculated as
\begin{widetext}
\begin{equation}
\frac{d\phi}{dt}=\frac{D_{eq}\bar{R}\bar{\gamma}_{C}\left(4D_{eq}+D_{eq}n_{2}^{2}-2n_{2}\right)\phi(t)}{4\bar{g}_{R}\bar{\gamma}_{C}\left(\tilde{D}D_{eq}+\tilde{D}n_{2}-2\right)+3\bar{\gamma}_{R}\left(8\tilde{D}D_{eq}+2\tilde{D}n_{2}-D_{eq}n_{2}\right)}.\label{EOMphi}
\end{equation}
The solution takes the form
\begin{equation}
\phi\left(t\right)=\phi_0e^{t/\tau},\label{Darkness}
\end{equation}
where $\phi_0$ is the initial phase angle of dark soliton. Here $\tau$ provides the characteristic time scale for the existence of solitons in presence of dissipation. It is explicitly given by
\begin{eqnarray}
\tau=\frac{3\bar{\gamma}_{R}}{\bar{R}\bar{\gamma}_{C}}f\left(\bar{g}_{R},n_{2}\right)=\frac{3}{\gamma_{C}\tau_{0}}\frac{P_{\text{th}}}{P-P_{\text{th}}}f\left(\bar{g}_{R},n_{2}\right),\label{DarkScale}
\end{eqnarray}
with
\begin{equation}
f\left(\bar{g}_{R},n_{2}\right)=\frac{8\tilde{D}D_{eq}+2\tilde{D}n_{2}-D_{eq}n_{2}+\left(2\bar{g}_{R}/3\bar{\gamma}_{R}\right)\left(\tilde{D}D_{eq}+\tilde{D}n_{2}-2\right)}{4D_{eq}^{2}+D_{eq}^{2}n_{2}^{2}-2n_{2}D_{eq}}.
\end{equation}
\end{widetext}
Notice that Eq. (\ref{DarkScale}) can be simplified into the expression
$\tau=3\bar{\gamma}_R/\bar{R}\bar{\gamma}_C$ as shown in Ref.~\cite{Smirnov2014} if $n_2=0$ and the term proportional to $\bar{g}_R$ is ignored. The knowledge of $\phi$ then allows us to determine the velocity  of the dark soliton via $v_{\text{s}}=D\tan\phi$ and hence the darkness of the dark soliton through $n^{\text{min}}_D=v^2_{\text{s}}(t)$.  It is immediately clear from Eq.~(\ref{Darkness}) that the dark soliton speeds up exponentially in time with a rate $\tau^{-1}$ until disappears eventually.

Equation (\ref{DarkScale}) shows that, also pointed out in Ref.~\cite{Smirnov2014}, the lifetime of soliton $\tau$ is proportional to $\bar{\gamma}_R$ while inversely proportional to $\bar{R}$ and $\bar{\gamma}_C$, as has been numerically verified.  In Fig.~\ref{Fig1} [see plots in the left and middle columns], we numerically solve Eqs.~(\ref{sss1}) and (\ref{sss2}) for the condensate density distribution of the dark soliton $|\psi_D(x,t )|^2$ and bright soliton $ |\psi_B(x,t )|^2$, taking $\bar{g}_R=2$, $\bar{\gamma}_C=3$ and $\phi_0=0.2$, while varying parameters $\bar{\gamma}_R$ and $\bar{R}$. We compare the numerical results for the darkness $n^{\text {min}}_1$ of the dark solitons with the analytical predictions from $n^{\text{min}}_D=v^2_{\text{s}}(t)$. A remarkable agreement between the two is found, as illustrated in the right column of Fig.~\ref{Fig1}. To be more specific, we compare Figs.~\ref{Fig1} (a3) and (b3),  corresponding to the cases with $\bar{R}=0.5$ and $\bar{R}=0.1$ by taking $\bar{\gamma}_R=30$, respectively. Whereas, comparisons of Figs.~\ref{Fig1} (a3) and (c3),  both taking by $\bar{R}=0.5$ while $\bar{\gamma}_R$ is different, demonstrates the effects of $\bar{\gamma}_R$. Obviously, the analytical results agree increasingly well with the numerical ones for larger $\bar{\gamma}_R$ and smaller $\bar{R}$. In all plots,  we note that the typical polariton relaxation time is about $\gamma^{-1}_C =10$ ps \cite{Rev1} and  the time scaling variable expressed as
$\tau_0 = \bar{\gamma}_C/\gamma_C$ with $\bar{\gamma}_C=3$ takes the physical value of 30 ps. The corresponding
propagation time for the solitonic state shown in Fig.~\ref{Fig1} reaches $t = 3000$ ps, which is much longer than the condensate and reservoir relaxation times.

Finally, we add the Langevin noise term Eq.~(\ref{Noise}) into Eqs.~(\ref{sss1}) and (\ref{sss2}), which serves both as an initial seed for the
condensate and to test the stability of the DB soliton against fluctuations. We compare the dynamics of dark soliton with [the third column of Fig.~\ref{Fig2}] and without noise [the first column of Fig.~\ref{Fig2}]. (The results for the bright soliton is compared in the second and fourth columns). We immediately see that, while noise in general leads to a faster decay of solitons, they remain stable until $t=20$, corresponding to the real time of $t=600 ps$. This means that the DB soliton can still propagate for a sufficiently long time even in presence of noise, hence making it possible to feasibly observe them in experiments.

\section{Discussion and Conclusion}\label{Conclusion}

As is known, a dark soliton in the DB soliton is more dynamically stable than a dark soliton alone. In the equilibrium case, it has been established that the dark soliton stripes in conservative atomic condensates and optical fields are always unstable against transverse excitations that have wavelength greater than their extension \cite{SnakeInstabilityReview,SnakeInstabilityFirst,SnakeInstability1}, leading to undulation and an eventual breakup of dark solitons into multivortex patterns \cite{SnakeInstabilityExpOpt,SnakeInstabilityExpBEC}.  However, the large DB solitons are expected to transcend this restriction, because their size can be much larger than their extension when the number of bright soliton becomes very large~\cite{DBexpBEC1}. 
In the non-equilibrium case, a natural question arises as to how the dissipative nature affecting the snake instability of DB solitons. Addressing this issue is beyond the scope of this work, and we will leave the snake instability of DB solitons for future investigations. 

In summary,  we have investigated the dynamics of dark-bright solitons appearing in spinor polariton Bose-Einstein condensates under non-resonant pumping. In particular, we have derived analytically the evolution equations for the soliton parameters. Within the framework of Hamiltonian approach, our analytical results capture the essential physics as to how the combined effects of the open-dissipative and spinor nature affects the dark-bright soliton by blending with the background at a finite time. We also solve the modified dissipative two-component GPEs in a numerically exact fashion. The numerical results find remarkable agreement with the analytically ones. We also demonstrate that these dissipative solitons can exist in a substantially long time  even in presence of noise, rendering therm available for experimental observations.

\begin{acknowledgments}
We thank Y. Xue, Ying Hu, and B. Wu for stimulating discussions. This is supported by the NSFC of China (Grants No. 11274315 and No. 11374125) and Youth Innovation Promotion Association CAS (Grant No. 2013125). L. C. is supported by the Science Foundation of Guizhou Science and Technology Department (GrantNo. QKHJZ [2017] 1202), and the Science Foundation of Guizhou Provincial Eduction Department (Grant No. QJHKYZ [2017] 087).  Z. D. Z. is supported by the NSFC of China (Grant No. 51331006).
\end{acknowledgments}

\bibliography{myr}
\end{document}